\documentclass[floatfix,a4paper,aps,prd,twocolumn,preprintnumbers]{revtex4-1}  
\usepackage{epsfig} 
\usepackage{axodraw} 
\usepackage{amsmath} 
\usepackage{amssymb} 
\usepackage{amsfonts} 
\usepackage{graphicx} 
\usepackage{graphics,subfigure} 
\usepackage{axodraw} 
\usepackage{float} 
\usepackage{booktabs} 
\usepackage{color} 
\usepackage{geometry} 
\usepackage{rotating} 
\usepackage{units} 
\usepackage{slashed} 
 
\newcommand{\bea}{\begin{eqnarray}} 
\newcommand{\eea}{\end{eqnarray}} 
\newcommand{\beq}{\begin{equation}} 
\newcommand{\eeq}{\end{equation}} 
\newcommand{\beqa}{\begin{eqnarray}} 
\newcommand{\eeqa}{\end{eqnarray}} 
\newcommand{\bit}{\begin{itemize}} 
\newcommand{\eit}{\end{itemize}} 
 
\newcommand{\st}{\ensuremath{\tilde{t}_1}} 
\newcommand{\neutr}{\ensuremath{\tilde{\chi}_1^0}} 
\newcommand{\charg}{\ensuremath{\tilde{\chi}_1^\pm}} 
\newcommand{\gev}[1]{\unit[#1]{GeV}} 
\newcommand{\mev}[1]{\unit[#1]{MeV}} 
\newcommand{\tev}[1]{\unit[#1]{TeV}} 
\newcommand{\met}{\ensuremath{\slashed{E}_T}} 
 
\newcommand{\ifb}[1]{\unit[#1]{fb^{-1}}} 

\newcommand{\Mtext}[1]{{\color{black} #1}}
 
\newbox\charbox 
\newbox\slabox 
\def\s#1{{      
    \setbox\charbox=\hbox{$#1$} 
    \setbox\slabox=\hbox{$/$} 
    \dimen\charbox=\ht\slabox 
    \advance\dimen\charbox by -\dp\slabox 
    \advance\dimen\charbox by -\ht\charbox 
    \advance\dimen\charbox by \dp\charbox 
    \divide\dimen\charbox by 2 
    \raise-\dimen\charbox\hbox to \wd\charbox{\hss/\hss} 
    \llap{$#1$} 
}}

\begin{document} 
  
\title{Light Stop Searches at the LHC with Monojet Events} 
 
\author{M. Drees} 
\email[]{drees@th.physik.uni-bonn.de} 
\affiliation{BCTP and Physics Institute, University of Bonn, Bonn, Germany } 
 
\author{M. Hanussek} 
\email[]{hanussek@th.physik.uni-bonn.de} 
\affiliation{BCTP and Physics Institute, University of Bonn, Bonn,  Germany} 
 
\author{J.~S. Kim} 
\email[]{jongsoo.kim@tu-dortmund.de} 
\affiliation{Institut f\"ur Physik, Technische Universit\"at Dortmund, D-44221
  Dortmund, Germany and ARC Centre of
  Excellence for Particle Physics at the Terascale, School of   Chemistry and Physics, University of Adelaide, Adelaide, Australia}

\begin{abstract} 
  We consider light top squarks (stops) in the minimal supersymmetric
  Standard Model at the Large Hadron Collider. Here, we assume that
  the lightest neutralino is the lightest supersymmetric particle
  (LSP) and the lighter stop is the next--to--LSP. Stop pair
  production is difficult to probe at the Large Hadron Collider for
  small stop--LSP mass splitting. It has been shown previously that
  even in this case stop detection is possible if one considers stop
  pair production in association with one hard jet. We reconsider this
  supersymmetric monojet signature and go beyond previous works by
  including the full Standard Model background and optimizing the
  cuts, working at the hadron level and including detector effects. As
  a result, a larger portion of the stop--LSP mass plane becomes
  accessible to monojet searches.
\end{abstract} 
 
\preprint{ADP11-40/T762} 
\preprint{DO-TH 12/03} 
 
\maketitle 
  
\section{Introduction} 
\label{sec:intro} 
 
Supersymmetric extensions of the Standard Model (SM) are popular among 
the large number of TeV--scale models \cite{susy}. In many 
supersymmetric models, the lightest supersymmetric particle (LSP) is 
stabilized by employing a discrete symmetry \cite{Ibanez:1991pr}. In 
the minimal supersymmetric extension of the SM (MSSM), the LSP is the 
lightest neutralino in large regions of parameter space. Since it 
escapes detection, the production of two  heavier superparticles and their 
subsequent decays into two LSPs plus several SM particles leads to the 
famous ``missing transverse energy ($E_T$)'' signature for supersymmetry. 
 
This signature has been searched for, most recently by the Large
Hadron Collider (LHC) experiments \cite{acms}. Unfortunately no signal
has yet been found. This allows to derive quite stringent bounds on
the masses of some strongly interacting superparticles. In particular,
first generation squarks and gluinos below about 1 TeV are excluded if
their masses are roughly equal.
 
This seems already somewhat high, considering that the main motivation 
for postulating the existence of superparticles is to stabilize the 
electroweak hierarchy against radiative corrections. However, to one 
loop order essentially only third generation (s)quarks appear in the 
loop corrections to Higgs mass parameters. Moreover, the analyses 
published by CMS and ATLAS so far are not sensitive to direct pair 
production of {\em only} third generation squarks, if the other 
squarks and gluinos are sufficiently heavy 
\cite{recent_stop}.\footnote{Very recently ATLAS published \cite{A_sb} 
  an analysis of a search for light sbottom pairs using about 2 
  fb$^{-1}$ of data, which excludes $\tilde b_1$ with mass below 400 
  GeV if $\tilde b_1$ decays with unit branching ratio into the 
  lightest neutralino, assuming the mass of that neutralino is 
  sufficiently small.} One reason is that the cross section for 
producing a pair of third generation squarks is much smaller than that 
for producing first generation squarks, since no ``flavor excitation'' 
contributions exist for third generation squarks. Hence stop masses of 
a few hundred GeV are still allowed, and in fact favored by finetuning 
arguments. 
 
In spite of their somewhat smaller production cross sections, the 
search for the pair production of third generation squarks at the LHC 
is in principle straightforward, as long as the mass difference to the 
LSP is sufficiently large. In this case one can employ the usual 
multi--jet plus missing $E_T$ (possibly plus one or more lepton(s)) 
signature; often demanding some of the jets to be tagged as $b-$jets 
will be helpful \cite{A_sb}. However, if the mass splitting to the 
LSP becomes small, all SM particles produced in stop decays will become 
soft, and the missing $E_T$ will therefore also become small.  The 
usual signals will then be swamped by background. 
 
At the same time, there are good reasons to assume that at least the
lighter stop mass eigenstate is significantly lighter than the other
squarks. First, if supersymmetry breaking is transmitted to the
visible sector at some high energy scale, Yukawa contributions to the
renormalization group evolution tend to reduce stop masses relative to
the masses of first generation squarks \cite{susy}. Furthermore,
mixing between the $SU(2)$ doublet left ($L-$)type and $SU(2)$
singlet right ($R-$)type squarks is proportional to the mass
of the corresponding quark, and is therefore most important for top
squarks. This mixing will further reduce the mass of the lighter
eigenstate (and increase that of the heavier eigenstate).
 
There are also more phenomenological reasons to be interested in quite 
light stops. One has to do with dark matter. As is well known, the 
lightest neutralino can be a viable dark matter candidate \cite{susy}, 
being weakly interacting and stable (if $R-$parity, or a similar 
symmetry, is exact).  However, for most combinations of parameters the 
computed LSP relic density is either too large (if the LSP is 
bino--like, which is preferred in many constrained models) or too 
small (if it is higgsino-- or wino--like). One (of several 
\cite{susy}) solutions is to have a bino--like neutralino with mass 
splitting of a few tens of GeV to the lightest stop. In this case 
co--annihilation \cite{coan} between these two states can lead to an 
acceptable relic density \cite{bdd}. 
 
Another reason to consider light top squarks is that in the context 
of the MSSM they are a necessary condition for electroweak (EW) 
baryogenesis \cite{Carena:2008vj,Li:2008ez,Huet:1995sh}. In fact, a 
MSSM scenario with a bino--like neutralino LSP and a light stop and 
(higgsino--like) chargino can simultaneously explain the observed 
baryon asymmetry in the universe and the cosmological dark matter. 
 
However, EW baryogenesis also requires a large CP violating phase in 
the chargino or neutralino sector. On the other hand, there are severe 
constraints on the CP phases necessary for EW baryogenesis from 
electron and neutron electric dipole moment (EDM) bounds 
\cite{susy_edm}. Thus, the phenomenological viable parameter space is 
strongly constrained. A simple solution is to assume that all other 
sfermions are very heavy \cite{Carena:2008vj,Nakamura:2010zzi}. In 
this case the (lighter) stop may well be the only squark that can be 
detected by the LHC experiments. 
 
Thus motivated, we study the effects at a hadron collider of a 
scenario where the lighter stop mass eigenstate $\tilde t_1$ is the 
only strongly interacting light sparticle, with rather small mass 
splitting to the neutralino LSP. We assume that charginos as well as 
all other neutralinos are heavier than $\tilde t_1$. The dominant 
sparticle production mechanism is then stop pair production. We 
further assume that the stop decay channels $\st\rightarrow b\neutr W$ 
and $\st\rightarrow t \neutr$ are kinematically closed and the 
four--body decay $\st\rightarrow \ell\nu_\ell b \neutr$ is strongly 
phase space suppressed, so that the loop induced two--body decay 
$\st\rightarrow c\neutr$ is the dominant decay mode 
\cite{Hikasa:1987db,abdel}. 
 
For sufficiently large mass splitting between the stop and the 
lightest neutralino, the charm jets become energetic enough to look 
for a di--jet plus missing $E_T$ signature of $\tilde t_1$ pair 
production. The Tevatron experiments were sensitive to mass splitting 
above $\gev{40}$ \cite{Abazov:2008rc,CDFnote9834}. Probing stop pair 
production in this channel becomes difficult for small mass splitting 
to the LSP, since the collider signature is then given by two very 
soft charms and correspondingly little transverse missing energy 
\cite{Balazs:2004bu}. However, including the effects of (both 
perturbative and non--pertubative) gluon radiation, $\tilde t_1$ pair 
production should be detectable at $e^+e^-$ colliders even for 
arbitrarily small mass splitting to the LSP \cite{de2}. In fact, LEP2 
experiments could rule out $\tilde t_1$ masses below $\gev{100}$ 
\cite{Achard:2003ge} even for very small mass splitting.  
 
One possibility to look for light $\tilde t_1$ nearly degenerate with 
the LSP at the LHC is via gluino pair production followed by $\tilde g
\rightarrow t \tilde t_1$ decays \cite{Kraml:2005kb}. At $\sqrt{s} =
\tev{14}$ the same--sign top signature should be detectable above the 
SM background for gluino masses up to $\gev{900}$. Another possibility 
\cite{Hiller:2009ii} is to look for a finite decay length of the light 
stop produced in gluino decays. 
 
In Ref.~\cite{Bornhauser:2010mw}, an alternative method to discover 
light stops in the co--annihilation region at the LHC was suggested, 
based on the associate production of a $\tilde t_1 \tilde t_1^*$ pair 
with a $b \bar b$ pair. For relatively light higgsinos, mixed 
electroweak (EW)--QCD contributions are large and can even exceed the 
pure QCD prediction, since there are $2 \rightarrow 3$ diagrams with 
an on--shell higgsino--like chargino decaying into a stop and a $b$ 
jet. These EW contributions are sensitive to the $\st-\charg-b$ 
coupling. Thus, they might be used to test a supersymmetry relation 
involving superpotential couplings, which has never been addressed so 
far. However, this requires that the masses of the lighter stop and 
the lightest neutralino are known, so that the pure QCD contribution, 
where the $b \bar b$ pair originates from gluon splitting, can be 
subtracted. Determining the stop mass from an independent, QCD 
dominated process would be advantageous for this purpose. Moreover, 
considering ${\cal O}(\alpha_S^3)$ processes might lead to a better 
discovery reach than the ${\cal O}(\alpha_S^2 \alpha_W, \,
\alpha_S^4)$ processes contributing to $\tilde t_1 \tilde t_1^+ b \bar
b$ production. 
 
Therefore, we here reconsider stop pair production in association with
a hard jet, which was proposed in Ref.~\cite{Carena:2008mj}. There it
was argued that the soft fragmentation and decay products of the stops
cannot be reconstructed as jets for small mass splitting. This
inevitably leads to the notion of a monojet \cite{Arnison:1984qu},
{\it i.e.} a final state containing a single high momentum jet, whose
$p_T$ is mostly balanced by the invisible LSPs, plus some soft
particles. In \cite{Vacavant} the SM background for monojets was
evaluated in the context of searching for extra dimensions; these
results were used in \cite{Carena:2008mj} to show that the monojet
signature from stop pair production can be seen above the SM
background up to stop masses of $\gev{200}$ or larger. Stop pair plus
photon production were also considered in \cite{Carena:2008mj}, but
due to the reduced cross section the mass reach of this channel is
even smaller.
 
In this work, we will re--analyze stop pair production in association 
with one hard jet. We perform a signal and background simulation at 
hadron level. We also simulate the most important detector effects by 
using a fast detector simulation. In \cite{Carena:2008mj} the selection 
cuts could not be optimized. Here, we develop a set of selection cuts 
optimized for searching for relatively light $\tilde t_1$ squarks 
nearly degenerate with the neutralino LSP. In addition, we also 
include $t\bar t$ as an important background for the monojet signal, 
which had been omitted in previous works.  
 
The remainder of this article is organized as follows. In 
Sect.~\ref{sec:signal_description} we describe our monojet signal. In 
Sect.~\ref{sec:lhc_analysis}, we first discuss the dominant background 
processes and then basic cuts for a benchmark scenario before 
presenting our numerical results. We show the discovery reach in the 
neutralino stop mass plane. We conclude in Sect.~\ref{sec:summary}. 
 
\section{Stop Pair Production in Association with a Jet at the LHC} 
\label{sec:signal_description} 
 
We consider stop pair production in association with one QCD jet,  
\beq \label{proc}
pp\rightarrow \st\st^* j + X,  
\eeq 
where $X$ stands for the rest of the event. We assume that the mass
difference between the lightest stop and the lightest neutralino is a
few tens of GeV or less, and that the on--shell $\tilde t_1
\rightarrow \tilde \chi_1^+ b$ and $\st \rightarrow b \neutr W$ decays
are closed. Due to the small mass splitting to the LSP, four--body
decays like $\tilde t_1 \rightarrow \tilde \chi_1^0 \ell^+ \nu_\ell b$
are strongly suppressed. However, the flavor changing neutral current
(FCNC) stop decay into a charm--quark and the lightest neutralino,
\beq \label{stdec}
\st\rightarrow c\neutr\,,
\eeq  
is open. This decay can only occur if $\tilde t_1$ has a 
non--vanishing $\tilde c$ component. As pointed out in 
\cite{Hikasa:1987db}, such a component will be induced radiatively 
through CKM mixing even if it is absent at tree level. For simplicity 
we assume that it has branching ratio of 100\%.  
 
The small mass difference to the LSP also implies that both charm 
``jets'' in the signal are rather soft.\footnote{Unless the stop 
  squarks themselves are highly boosted, which is true only in a tiny 
  fraction of all signal events.} The charm quarks will then not be 
useful for suppressing backgrounds since soft jets are ubiquitous at 
the LHC, and may not be detected as jets at all. Thus our signal will 
be a single high $p_T$ jet with large missing energy, 
\beq \label{signal}
pp\rightarrow j\met, 
\eeq  
possibly accompanied by one or more soft jet(s) from gluon radiation
and the $\tilde t_1$ decay products. At the LHC, the largest
contribution to stop pair production in association with a jet comes
from gluon fusion diagrams, but contributions from $qg$ and $\bar q g$
initial states, which become more important for large $\tilde t_1$
masses, are nearly as large.\footnote{For light stop masses of
  \gev{120}, the $qg$ contribution is already about $43$\% of the
  total cross section. It increases to $47\%$ for \gev{300} stops.}
Contributions from $q \bar q$ annihilation are relatively small. We
perform a full leading order analysis, using exact ${\cal
    O}(\alpha_S^3)$ parton--level cross sections for $gg, \, q \bar q
  \rightarrow \tilde t_1 \tilde t_1^* g$ and $g q \rightarrow \tilde
  t_1 \tilde t_1^* q$.
 
Since most events have at least one gluon in the initial state, we 
expect strong QCD bremsstrahlung due to the large color charge. The 
QCD radiation increases with increasing stop mass. However, the 
topology of the signal is still simple compared to standard 
supersymmetric collider signatures: a single energetic jet, which is 
essentially back to back to the missing transverse momentum vector. 
 
\section{Numerical Analysis} 
\label{sec:lhc_analysis} 
 
In this section, we discuss the major backgrounds, and describe how to
determine them from experimental data, including a discussion of the
resulting systematic and statistical errors. Next, we shortly discuss
our numerical tools before introducing a specific benchmark
scenario. We then show the relevant kinematic distributions
and motivate our final cuts. We conclude the section with the
discovery reach at the LHC in the neutralino stop mass plane.
 
\subsection{Backgrounds}\label{subsec:backgrounds}

\begin{table*}[t] 
\begin{center} 
\begin{ruledtabular} 
\begin{tabular}{c|cccc}
process & $Z(\rightarrow \nu\bar\nu)+j$ & 
$W(\rightarrow e\nu_e,\mu\nu_\mu)+j$ & $W(\rightarrow\tau\nu_\tau)+j$
& $t\bar t$ \\ \hline \
$\sigma$ [pb] & 37 & 94 & 47 & 800 \\
\end{tabular}
\end{ruledtabular} 
\caption{Total hadronic cross sections in pb for the main SM
  backgrounds at $\sqrt{s}=14$ TeV. The cross sections were calculated 
  with {\tt Pythia8.150} apart from $t\bar t$ production, which is 
  calculated in Ref.~\cite{Bonciani:1998vc}. The $V+j \ (V = W, Z)$ 
  cross sections have been calculated demanding $p_T > \gev{150}$ for 
  the parton--level jets.
  \label{tab:bkg_xs}}
\end{center} 
\end{table*}

The dominant SM backgrounds are: 
 
\bit 
 
\item $Z(\rightarrow \nu \bar\nu)+j$ production, {\it i.e.} $Z$ boson 
  production in association with a jet. The $Z$ boson decays into a 
  pair of neutrinos. If the charm jets in the signal are very soft, 
  this background looks very similar to our signal.  We will see in 
  Section \ref{sec:discovery_potential} that $Z(\rightarrow \nu 
  \bar\nu)+j$ is the dominant irreducible background after applying 
  all kinematic cuts. Fortunately its size can be directly determined 
  from data: One can measure $Z(\rightarrow \ell^+ \ell^-)+j$, where 
  the $Z$ decays into a pair of either electrons or muons.  From the 
  known $Z$ branching ratios (BRs) one can then obtain an estimate for 
  the background cross section. However, this procedure will increase 
  the statistical error, since $BR(Z \rightarrow \ell^+ \ell^-) \simeq 
  BR(Z \rightarrow \nu_i \bar \nu_i) / 3$ after summing over $\ell = 
  e, \, \mu$ and all three generations of neutrinos. Moreover, not all 
  $Z \rightarrow \ell^+ \ell^-$ events are reconstructed 
  correctly. Including efficiencies, Ref.~\cite{Vacavant} estimated 
  that the calibration sample $Z(\rightarrow e^+e^-/\mu^+\mu^-)+j$ is 
  roughly a factor of $5.3$ smaller than the $Z(\rightarrow \nu 
  \nu)+j$ background in the signal 
  region.\footnote{Ref.~\cite{Vacavant} cites a factor of seven 
    between the $Z \rightarrow \ell^+ \ell^-$ control sample and the 
    {\em total} background from $V+j$ production ($V = W^\pm, Z$), 
    after applying a lepton veto in the signal. The ratio of $5.3$ 
    follows since according to the cuts of \cite{Vacavant}, about 75\% 
    of the $V+j$ background comes from $Z(\rightarrow \nu \bar \nu) + 
    j$.}  Hence, we expect that the error of this background is 
  $\sqrt{5.3} \simeq 2.3$ times larger than the statistical error. 
 
\item $W(\rightarrow \ell \nu) + j$ production, where the $W$ decays 
  leptonically. Unlike the signal, this background contains a charged 
  lepton ($\ell=e^\pm,\mu^\pm$), and will thus resemble the signal 
  only if the charged lepton is not identified. This can happen when 
  the charged lepton emerges too close to the beam pipe or (in case of 
  electrons) close to a jet. Since the production cross section for 
  $W(\rightarrow \ell \nu)+j$ is larger than $Z(\rightarrow \nu \bar 
  \nu)+j$ by a factor of $\sim 3$, this will still contribute 
  significantly to the overall background, as we will see in Section 
  \ref{sec:discovery_potential}. The $W(\rightarrow \ell \nu_\ell) +j$ 
  background can be determined by extrapolation using events where the 
  lepton is detected. 
 
\item $W(\rightarrow \tau \nu) + j$ production, where the $W$ decays 
  into a tau. The reconstructed jets from a hadronically decaying tau 
  are in general not back to back in azimuth to the missing momentum 
  vector. Ref.~\cite{Vacavant} exploits this feature to suppress the 
  tau decay channel of $W+j$. However, identification of hadronically 
  decaying $\tau$ leptons is not easy. This background can be 
  experimentally determined with the help of $W(\rightarrow \ell \nu) 
  + j$ events where the charged lepton is detected, using known tau 
  decay properties. We (quite conservatively) assign an overall 
  systematic uncertainty of $10\%$ for the total $W+j$ background, 
  including that from $W \rightarrow \ell \nu_\ell$ decays. 
 
\item $t\bar t$ production (including all top decay channels). Top 
  decays will almost always produce two $b-$jets. Since we require 
  large missing $E_T$, at least one of the $W$ bosons produced in top 
  decay will have to decay leptonically. Note that this again gives 
  rise to a charged lepton ($e, \mu$) or $\tau$, whereas the signal 
  does not contain isolated charged leptons. However, for hadronically 
  decaying $\tau$'s, we can have large missing $E_T$ with no $e$ or 
  $\mu$ present. This background can again be estimated by normalizing 
  to $t \bar t$ events where (at least) one charged lepton is 
  detected. Just as for the $W+j$ background, we assume a total 
  systematic error of $10\%$. 
 
\eit 
 
We consider the above default estimates of systematic errors
to be conservative, since they do not rely on Monte Carlo
simulations. We expect that the SM contribution to the missing
transverse energy signal rate will be determined with at least this
precision. For example $W^\pm + j$ and even $\gamma + j$ samples can
also be used for reducing the error on the leading $Z( \rightarrow \nu
\bar \nu) + j$ background, since these classes of events have very
similar QCD dynamics \cite{vj_bckgd}.
 
In principle, one should also consider single top production, since
semi--leptonic top decays can again give rise to large missing
$E_T$. However, the production cross section for single top production
is a factor of $\sim \!4$ smaller than for $t \bar t$. Even though $t
\bar t$ is important for the cut selection, we will see that in the
end it only contributes $5\%$ to the total SM background. For these
reasons, we neglect single top production as a background.  We do not
consider pure QCD dijet and trijet production in our analysis, since a
large $\met$ cut is expected to essentially remove those backgrounds
\cite{Aad:2009wy, atlasConf, Allanach:2010pp}. We also neglect gauge
boson pair production as background, since the total cross section is
much smaller than that for single gauge boson plus jet production.
 
There are many SUSY processes leading to a monojet signature, which 
could be considered to be backgrounds to our signal. LSP pair plus jet 
production always gives a monojet signature, but has a very small 
cross section. Associate gluino plus squark production can lead to 
monojets if the gluino mass is close to that of the neutralino LSP. In 
addition, squark pair production can give rise to monojets, if both 
squarks directly decay into the LSP and one of the two jets is lost in 
the beam direction or the partons from both squarks are reconstructed 
in the same jet. Recently, \cite{Allanach:2010pp} considered 
squark--wino production. However, as we argued in the introduction, in 
order to avoid bounds from electron and neutron EDM, we assume that 
most superparticles are quite heavy. Thus, the production rates of 
these additional supersymmetric processes are strongly suppressed and 
we need only consider Standard Model backgrounds. 
 
Estimates for the total hadronic cross sections for these SM
backgrounds are given in Table~\ref{tab:bkg_xs}. The cross section for
the $t \bar t$ background has been taken from \cite{Bonciani:1998vc},
which includes NLO corrections as well as resummation of
next--to--leading threshold logarithms.  The $W, \, Z +$ jet
backgrounds have been calculated with {\tt
  Pythia8.150}~\cite{Sjostrand:2007gs}. Note that exact ${\cal
  O}(\alpha \, \alpha_S)$ parton--level cross sections have been used
to generate the hardest jet in the $W, \, Z +$ jet backgrounds. We
checked explicitly that using also exact matrix element for the
emission of the second jet and matching to the parton shower does not
change the final background estimate (see below).
 
We have generated $2\cdot 10^6$ $Z(\rightarrow \nu \bar \nu)+j$
events, $2\cdot10^6$ $W(\rightarrow e\nu_e,\mu\nu_\mu)+j$ events,
$2\cdot10^6$ $W(\rightarrow\tau\nu_\tau) +j$ events as well as $10^7$
$t\bar t$ events.

\subsection{Numerical tools} \label{sec:Numerical_tools}
 
The masses, couplings and branching ratios of the relevant sparticles
are calculated with {\tt SPheno2.2.3} \cite{Porod:2003um}, starting
from weak--scale inputs for the relevant parameters. We use the
CTEQ6L1 parton distribution functions and the one--loop expression for
the strong gauge coupling with five active flavors with
$\Lambda_{\rm{QCD}}=\mev{165}$ \cite{Pumplin:2002vw}. Our
parton--level signal events are generated with {\tt Madgraph4.4.5}
\cite{Maltoni:2002qb}. These events are then passed on to {\tt
  Pythia8.150} \cite{Sjostrand:2007gs} for showering and
hadronization.  As already mentioned, we generate our SM background
events directly with {\tt Pythia8.150} fixing the $t\bar t$
normalization as in Table~\ref{tab:bkg_xs}. Except for the $t\bar t$
sample, we employed a parton--level cut of $\gev{150}$ on the minimum
transverse momentum of our parton--level jet, which will become the
``monojet'' in our signal and background; the final cut on the $p_T$
of this jet will be much harder, so that the cut on the parton--level
jet, which greatly increases the efficiency of generating signal and
background events, does not affect our final results. 

\Mtext{
In our default treatment only the emission of the first jet in the
signal process (\ref{proc}) as well as in the $(W,Z) +$ jet
backgrounds has been treated using exact matrix elements. The parton
shower is then allowed to produce additional jets whose hardness is
limited only by the initial and final state shower scales, for which
we adopt the default values. It is known that this does not describe
the emission of additional very energetic jets accurately. However, we
will see later that we have to reject events containing a second hard
jet, and the emission of relatively soft partons should indeed be
described adequately by the shower algorithms. In case of the $Z+$
jet background we checked explicitly that this is indeed the case, by
matching $Z + 2$ jet production described by exact matrix elements
with the parton shower, using the MLM matching algorithm \cite{mlm};
see Appendix A for details.

In case of the $t \bar t$ background, only the exact leading order
matrix element for the $2 \rightarrow 2$ process has been employed,
together with exact matrix elements describing top decay. Since $t
\bar t$ decay by itself produces between 2 and 6 hard partons, there
is no need to model the emission of additional hard partons, which
will be vetoed anyway, using exact matrix elements. We checked
explicitly that $t \bar t j$ events generated using exact matrix
elements for the emission of the additional parton $j$ with $p_T(j) >
150$ GeV at parton level, have an even smaller acceptance after cuts
(outlined below) than simple $t \bar t$ events.

Finally, in case of the signal additional jet activity can originate
from the charm quarks produced in $\tilde t_1$ decays. However, these
jets have basically no chance to pass our very stiff cut on the $p_T$
of the hardest jet. We therefore did not generate simple $\tilde t_1
\tilde t_1^*$ pairs without additional parton in the final
state. Since our signal estimate is exclusively based on the reaction
(\ref{proc}), issues of double counting do not arise.}

Our events are stored in the Monte Carlo event record format {\tt
  HepMC 2.04.01} \cite{Dobbs:2001ck}.  We take into account detector
effects by using the detector simulation {\tt Delphes1.9}
\cite{Ovyn:2009tx}, where we choose the default ATLAS--like detector
settings. Our event samples are then analyzed with the program package
{\tt ROOT} \cite{Brun:1997pa}.
 
\begin{table*}[t!] 
\begin{center} 
\begin{ruledtabular} 
\begin{tabular}{c|ccccccccccc} 
$m_{\st}$ [GeV] & 120    & 140    & 160   & 180 & 200 & 220 & 240 & 
260 & 280 & 300 & 320 \\ 
\hline 
$\sigma$ [pb] & 31 & 20 & 13 & 8.8 & 6.0 & 4.2 & 2.9 & 2.1 & 1.5 & 1.2 
& 0.86\\ 
\end{tabular} 
\end{ruledtabular} 
\caption{Total hadronic cross sections in pb for the signal at 
  $\sqrt{s}=14$ TeV. The cross sections were calculated with {\tt 
    Madgraph4.5.5}, with a parton--level cut $p_T > \gev{150}$ on the 
  jet. 
\label{tab:signal_xs}} 
\end{center} 
\end{table*} 
 
We define jets using the anti$-k_t$ algorithm implemented in FastJet
\cite{Cacciari:2006sm}, with a cone radius $\Delta R = \sqrt
{(\Delta\phi)^2 + (\Delta\eta)^2} = 0.7$ , where $\Delta \phi$ and
$\Delta\eta$ are the difference in azimuthal angle and rapidity,
respectively. All jets have to have $p_T > \gev{20}$. We demand that
electrons have $p_T(e) > \gev{10}$ and are isolated, i.e. that there
is no other charged particle with $p_T > \gev{2.0}$ within a
cone radius $\Delta R = 0.5$. Since muons can be identified even if
they are not isolated and have quite small $p_T$
\cite{Franchino:2011pv}, we include all reconstructed muons with $p_T
> \gev{4}$. Note that {\tt Delphes1.9} assumes a track reconstruction
efficiency of only 90\%, giving a substantial probability that charged
leptons are lost. Moreover, we only include true leptons, i.e. we do
not attempt to estimate the rate of fake leptons.
 
In {\tt Delphes1.9}, the same object can in principle be reconstructed 
as several different objects. {\it E.g.}, an electron can be 
reconstructed as an electron as well as a jet. Since such double 
counting of objects has to be prevented, we use an object removal 
procedure similar to that outlined in Ref.~\cite{Aad:2009wy}. However, 
any jet within $\Delta{\rm R}<0.2$ of an electron (including 
non--isolated electrons) will be removed if 
\beq 
p_T({\rm jet})-p_T(e^\pm)<\gev{20}\,. 
\eeq 
This removes ``jets'' whose energy is dominated by an electron, but we 
keep hard, hadronic jets even if they are very close to an 
electron. Note that contrary to Ref.~\cite{Aad:2009wy}, we keep all 
isolated electrons and all muons even if they are close to a jet. 
  
\subsection{Benchmark Scenario} 
\label{sec:benchmark_sceanrio} 
 
In the introduction, we motivated scenarios with a light stop and
light neutralino in order to be fully consistent with dark matter and
electroweak baryogenesis. However, in this study we do not only want
to discuss these scenarios, but also to determine the discovery reach
in the stop--neutralino plane, where the mass difference between stop
and lightest neutralino is at most a few tens of GeV. On the one hand,
scenarios with a heavier stop are expected to have a worse signal to
background ratio than those with a very light stop, due to the very
quickly decreasing production cross section. However, for heavier
stops, producing an additional hard jet reduces the cross section by a
smaller factor than for light stops. We choose a scenario with a
rather large stop mass, in order to probe the discovery reach found in
Ref.~\cite{Carena:2008mj}:
\beq 
m_{\st}=\gev{220}, 
\label{eq:mstop} 
\eeq 
as a benchmark scenario. The mass of the lightest neutralino 
is 
\beq 
m_{\neutr}=\gev{210}. 
\label{eq:mneut} 
\eeq 
All remaining sparticles are decoupled.\footnote{In order to reduce 
  stop and sbottom loop contributions to electroweak precision 
  variables, in particular to the $\rho$ parameter \cite{rho}, our 
  $\tilde t_1$ should be predominantly an $SU(2)$ singlet. However, 
  the stop mixing angle and the identity of the LSP are irrelevant for 
  our analysis. Similarly, the presence of relative light 
  higgsino--like chargino and neutralino states, as required for EW 
  baryogenesis, does not affect our ana\-ly\-sis, as long as they are not 
  produced in $\tilde t_1$ decays.} 
 
We require $p_T({\rm jet}) \ge \gev{150}$ for the parton--level jet. 
The total leading order (LO) cross section for our signal then only 
depends on the stop mass. The cross section for the benchmark point is 
$\sigma=4.2$ pb. We have generated $8\cdot10^5$ signal events for our 
benchmark point. LO predictions for cross sections for different stop 
masses are listed in Table~\ref{tab:signal_xs}. 
 
As described in Sect.~\ref{sec:signal_description} we assume that all 
$\tilde t_1$ undergo two--body decay 
\beq 
\st\rightarrow c\neutr\,. 
\eeq 
We assume that these decays are prompt; a finite impact parameter 
would greatly facilitate detection of the signal \cite{Hiller:2009ii}.

\subsection{Distributions}\label{sec:distributions} 
 
In this Subsection, we discuss the basic kinematic distributions and 
jet and particle multiplicities for the signal as well as for the 
background processes. The distributions are not stacked on each other 
and are shown on a logarithmic scale. All distributions are scaled to 
an integrated luminosity of $\ifb{100}$ at $\sqrt{s}=\tev{14}$ at the 
LHC. 
 
We show in Fig.~\ref{fig:nlepton_distribution} the number of leptons
(electrons and muons, as defined in Subsect.~B) for signal and
background. The signal contains very few charged leptons. In
principle, semi--leptonic $c \rightarrow s \ell \nu_\ell$ decays can
produce leptons, but these are usually too soft to sa\-tis\-fy our
criteria; in addition, most of the remaining electrons are removed by
our isolation criterion. The $Z+j$ background also contains very few
leptons, since we only consider $Z\rightarrow\nu\bar\nu$ decays here.
In contrast, the $t\bar t$ background can have up to seven charged
leptons, mostly from semileptonic $t \rightarrow b \rightarrow c
\rightarrow s,d$ decays. Note that we include $t \bar t$ events where
both $t$ quarks decay fully hadronically. This background therefore
peaks at $n_\ell = 0$ charged leptons. The $W+j$ background peaks at
$n_\ell = 1$ charged lepton; recall that we have only generated
$W\rightarrow\ell\nu$ decays here, and that we show the $W+j$
background with $W\rightarrow \tau\nu_\tau$ separately. In the latter
case a charged lepton can arise from the leptonic decays of the tau.
 
\begin{figure} 
\vspace{+1.5mm} 
\includegraphics[width=0.3\textwidth,angle=-90]{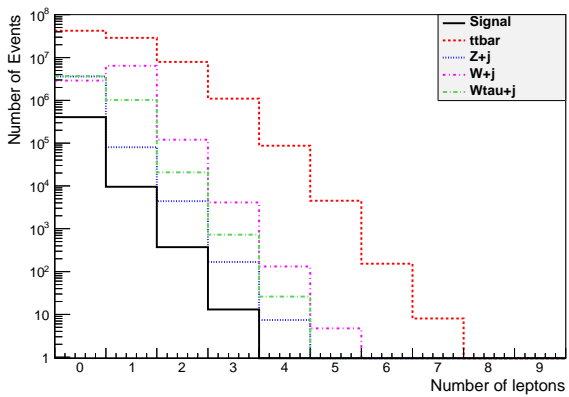} 
\caption{Number of leptons for the signal and SM backgrounds 
  assuming an integrated luminosity of $100\,\text{fb}^{-1}$ at 
  $\sqrt{s}=14$ TeV.  For the signal we assumed the benchmark scenario 
  of Sect.~\ref{sec:benchmark_sceanrio}, {\it i.e.} 
  $m_{\tilde{\chi}_1^0} = \gev{210}$ and $m_{\st} = 
  \gev{220}$.} 
\label{fig:nlepton_distribution} 
\end{figure} 
 
We will later apply a hard cut on missing $E_T$. This would remove all 
$W+j$ events where the $W$ decays hadronically, which we therefore 
didn't bother to generate. Similarly, $t \bar t$ events can pass this 
cut only if they contain at least one charged lepton.\footnote{Since 
  the other top (anti)quark might decay fully hadronically, we cannot 
  simply enforce semi--leptonic top decays when simulating this 
  background.}  A veto on charged leptons will therefore efficiently 
remove most of the SM backgrounds, except for the contribution from 
$Z(\rightarrow \nu \bar \nu) +j$. 
 
The distribution of the number of identified taus is shown in
Fig.~\ref{fig:ntau_distribution}. Leptonically decaying taus cannot be
reconstructed; they can, however, be vetoed by charged lepton veto, if
the decay lepton is sufficiently energetic. On the other hand, taus
decaying hadronically {\em can} be identified, although tau
identification is not very easy at a hadron collider. In case of
hadronic tau--decays, only 1--prong events are taken into account for
the reconstruction of tau--jets in {\tt Delphes}, where $77\%$ of all
hadronically decaying taus are 1--prong events. {\tt Delphes} exploits
that the cone of tau jets is narrower than that of QCD jets and they
state a tau--tagging efficiency of about $30\%$ for $Z \rightarrow
\tau^+ \tau^-$. We find that the tau tagging efficiency, as estimated
by {\tt Delphes}, is much worse for the $t\bar t$ background due to
the increased hadronic activity. Nevertheless the $t\bar t$ background
has the second largest percentage of identified taus, exceeded only by
$W(\rightarrow \tau\nu)+j$; even in the latter case only about 25\% of
all events contain an identified tau, even though {\em all} of these
events do contain a tau lepton.\footnote{It might well be possible to
  design a tau {\em veto} algorithm that performs better than that
  used by {\tt Delphes}. We have not attempted to do so since at the
  end the SM background will be dominated by $Z \rightarrow \nu \bar
  \nu$ events even assuming {\tt Delphes} efficiencies.} Note that we
include mis--tags of QCD jets as taus, as estimated by {\tt
  Delphes}. In fact, most $\tau-$jets identified in the signal are
fakes.

\begin{figure} 
\vspace{+1.5mm} 
\includegraphics[width=0.3\textwidth,angle=-90]{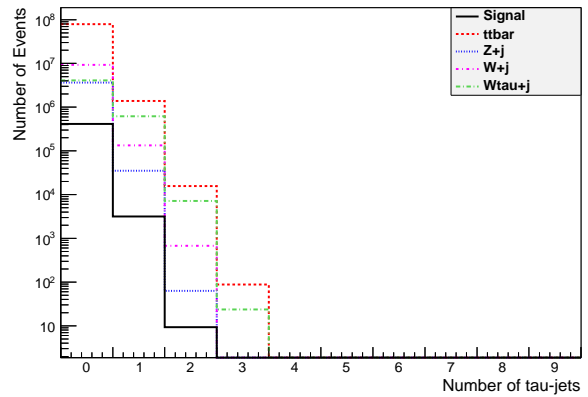} 
\caption{Number of isolated hadronic taus for the signal and SM 
  backgrounds. Parameters are as in Fig.~1.} 
\label{fig:ntau_distribution} 
\end{figure}

Fig.~\ref{fig:njet_distribution} shows the number of reconstructed
jets including $b-$jets. Jets are reconstructed with the anti$-k_t$
jet algorithm with a cone of $\Delta R=0.7$. We require the jets to
have minimum transverse momentum $p_T>\gev{20}$. We see that the
signal distribution has its peak around four jets. Jets can be created
not only from the hard interaction (e.g. the jet produced explicitly
in the signal as well as in the $V+j$ backgrounds, or the jets
produced in top decays), but also from QCD radiation in the initial
and/or final state. QCD radiation is controlled by the average
partonic squared center of mass energy $\hat s$ as well as by the
color charges in the initial and final states. As expected from the
discussion in Section \ref{sec:signal_description}, we see that the
jet multiplicity of the signal is on average larger than for the gauge
boson plus jet backgrounds. Not surprisingly, the $t \bar t$
background is characterized by the by far largest average jet
multiplicity. In previous works, the $t\bar t$ background was
omitted. Fig.~\ref{fig:njet_distribution} indicates that this
background can be greatly reduced by cutting against additional jet
activity; however, such a cut will reduce the signal more than the
$V+j$ backgrounds. Therefore, it is crucial to include the $t \bar t$
background in our analysis in order to determine the optimal set of
cuts.

\begin{figure} 
\vspace{+1.5mm} 
\includegraphics[width=0.3\textwidth,angle=-90]{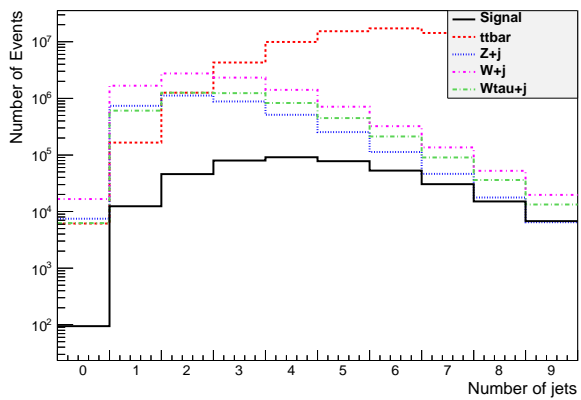} 
\caption{Number of jets for the signal and SM backgrounds. Parameters 
  are as in Fig.~1.} 
\label{fig:njet_distribution} 
\end{figure}

Fig.~\ref{fig:nbjets_distribution} shows the number of tagged 
$b-$jets. A jet is taggable as a $b-$jet if it lies in the acceptance 
region of the tracking system, i.e. satisfies $|\eta| < 2.5$ in 
addition to the requirement $p_T > \gev{20}$ that all jets have to 
fulfill, and if it is associated with the parent $b-$quark. {\tt 
  Delphes} assumes a tagging efficiency of about $40\%$ for taggable 
jets; the total $b-$tagging efficiency is thus less than $40\%$. {\tt 
  Delphes} also assumes mistagging efficiencies of $10\%$ and $1\%$ 
for charm--jets and light--flavored (or gluon) jets, respectively. Not 
surprisingly, the $t\bar t$ background contains the largest number of 
$b-$tags, since every $t\bar t$ event contains two $b-$quarks arising 
from top quark decays, and additional $b-$quarks can emerge from gluon 
splitting. Unfortunately the signal contains $b-$tags slightly more 
often than the $V+j$ backgrounds do. This is partly due to the 
presence of two $c$ (anti)quarks, which have a relatively high 
probability to be mistagged as $b-$jets. Moreover, at the 
parton--level the jet in signal events is most often a gluon, which 
can split into a $b \bar b$ pair, whereas in $V+j$ events the 
parton--level jet is most of the time a quark; signal events are 
therefore more likely to produce a $b \bar b$ pair in the QCD shower. 
Nevertheless a $b-$jet veto will suppress the $t\bar t$ background 
with relatively little loss of signal.

\begin{figure} 
\vspace{+1.5mm} 
\includegraphics[width=0.3\textwidth,angle=-90]{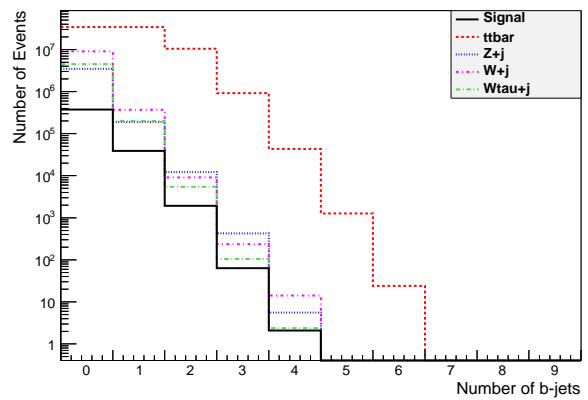} 
\caption{Number of tagged $b-$jets for the signal and SM 
  backgrounds. Parameters are as in Fig.~1.} 
\label{fig:nbjets_distribution} 
\end{figure}

The $p_T$ distribution of the hardest jet is given in 
Fig.~\ref{fig:jetpt1_distribution}, where we have also included the 
$b-$jets. At very large transverse momentum, $p_T({\rm jet}) > 
\gev{600}$, all curves have similar slopes, since then the hardness of 
the event is determined by the $p_T$ of the hardest jet rather than 
the mass of the produced particles. However, at smaller $p_T$ the 
$V+j$ backgrounds have a significantly softer spectrum than the signal 
and the $t \bar t$ background; once a pair of massive particles is 
produced, producing a jet with $p_T$ comparable to, or smaller than, 
twice the mass of these particles is more likely than in events 
containing only relatively light particles. Finally, the peaks in the 
distributions for the signal as well as the $V+j$ backgrounds are due 
to the parton--level cut of $\gev{150}$ on the jet that is produced as 
part of the hard partonic collision. Recall that we generated $t\bar 
t$ events without requiring an additional parton, and therefore we did 
not require a minimum $p_T$(jet1) here at parton--level. As a result, 
the $t\bar t$ contribution peaks at a lower $p_T$ value ($\sim m_t/2$, 
off the scale shown in Fig.~\ref{fig:jetpt1_distribution}) than the 
other processes. We conclude from Fig.~\ref{fig:jetpt1_distribution} 
that a lower cut of about $\gev{500}$ on the hardest jet will improve 
the statistical significance of the signal.

\begin{figure} 
\vspace{+1.5mm} 
\includegraphics[width=0.3\textwidth,angle=-90]{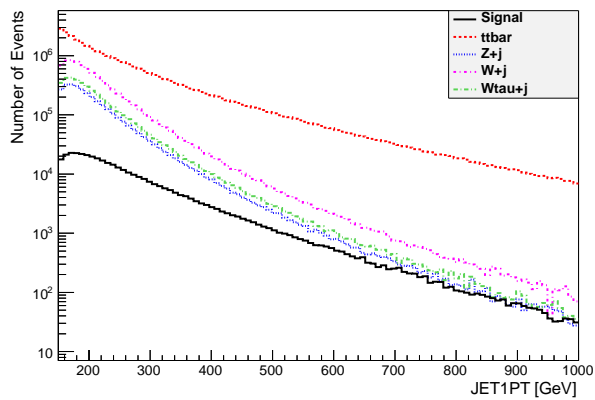} 
\caption{$p_T$ distributions of the hardest jet for the signal and SM 
  backgrounds. Parameters are as in Fig.~1.} 
\label{fig:jetpt1_distribution} 
\end{figure}

We see in Fig.~\ref{fig:jetpt2_distribution} that the $p_T$ 
distribution of the second hardest jet is much softer for the signal 
and the $V+j$ backgrounds than that of the hardest jet. Recall that  
the first jet is generated at parton level with $p_T > \gev{150}$, whereas 
the second jet comes from QCD showers, or, in case of the signal, 
possibly from stop decays; both sources give mostly soft jets, whose 
spectrum is backed up against the lower cut of $\gev{20}$ we impose on 
all jets. In contrast, in $t \bar t$ events the hardest and second 
hardest jet usually both originate from top decay. The $p_T$ spectrum 
of the second hardest jet therefore peaks not much below that of the 
hardest jet, at $p_T \simeq \gev{75}$.

\begin{figure} 
\vspace{+1.5mm} 
\includegraphics[width=0.3\textwidth,angle=-90]{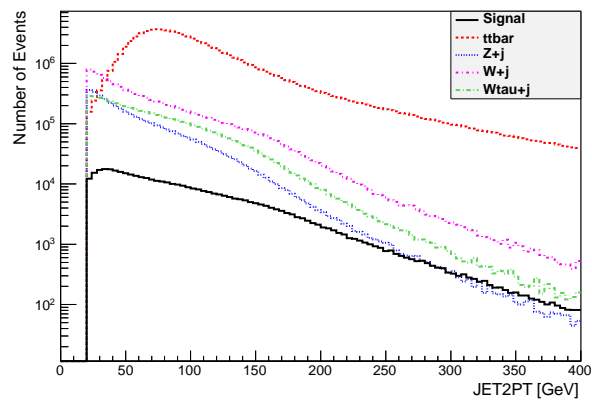} 
\caption{$p_T$ distributions of the second hardest jet for the signal 
  and SM backgrounds. Parameters are as in Fig.~1.} 
\label{fig:jetpt2_distribution} 
\end{figure}

From Fig.~\ref{fig:njet_distribution}, we have seen that a veto on the 
second jet is necessary in order to sufficiently suppress the $t\bar 
t$ background. However, if we vetoed all jets with $p_T>\gev{20}$, we 
would lose too many signal events. We find that it is a good choice to 
veto all events where the second hardest jet has $p_T> \gev{100}$. We 
also examined a veto on the third hardest jet with reduced $p_T$ 
threshold. This would reduce the $t\bar t$ background even 
further. However, it would also remove many signal events and thus a 
veto on the third jet does not increase the significance of our signal. 
 
Finally, Fig.~\ref{fig:met_distribution} shows the missing transverse 
energy distributions of signal and backgrounds. We see that the signal 
has the slowest fall off. Recall that we did not take into account 
pure QCD backgrounds such as dijet and trijet events. Thus we need a 
cut on missing energy in order to suppress these backgrounds 
\cite{Aad:2009wy}. We find that a missing transverse energy cut near 
$\gev{450}$ maximizes the significance of the signal for our benchmark 
point. Such a hard cut on the missing $E_T$, together with the veto on 
a second hard jet, should suppress the pure QCD background to a 
negligible level.

\begin{figure} 
\vspace{+1.5mm} 
\includegraphics[width=0.3\textwidth,angle=-90]{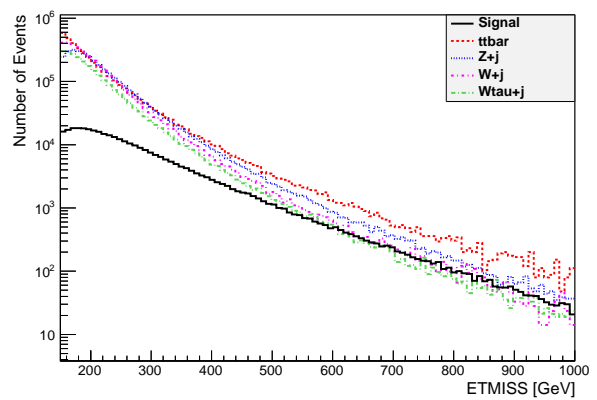} 
\caption{Missing transverse energy distributions for the signal and SM 
  backgrounds. Parameters are as in Fig.~1.} 
\label{fig:met_distribution} 
\end{figure}

\subsection{Discovery Potential at the LHC}  
\label{sec:discovery_potential}  
 
In the previous Subsection, we have discussed the basic distributions 
which we use to derive a set of kinematical cuts. Now we discuss the 
statistical significance for our benchmark point. Then, we will show 
the discovery potential of our signal in the stop--neutralino mass 
plane at the LHC for an integrated luminosity of $\ifb{100}$ at 
$\sqrt{s}=\tev{14}$, using the same set of cuts that optimizes the 
signal significance for our benchmark point.  
 
As motivated by our discussion in subsection \ref{sec:distributions}, 
we apply the following set of cuts: 
 
\bit 
\item $p_T(\rm{jet}_1) \geq \gev{500}$, i.e. we require one hard jet 
  with $p_T\geq\gev{500}$.  
\item $\met>\gev{450}$, i.e. we demand large missing transverse energy. 
\item $N_{\rm{lepton}}<1$, i.e. we veto all events with a 
  reconstructed electron or muon with $|\eta|<2.5$. Recall that we 
  only include isolated electrons with $p_T>\gev{10}$, but all muons 
  with $p_T > \gev{4}$. 
\item $N_{\rm{tau}}<1$, i.e. we veto all events with an identified tau 
  jet with $|\eta|<2.5$ and $p_T>\gev{20}$. 
\item $N_{\rm{b-jet}} <1$, i.e. require a veto on all tagged $b-$jets 
  with $p_T>\gev{20}$ and $|\eta|<2.5$. 
\item $p_T(\rm{jet}_2)<\gev{100}$, i.e. we veto the existence of a
  second hard jet.  
\eit 
The numerical values of the first, second and last cut have been set
by optimizing the signal significance for our benchmark point.

In Table \ref{tab:cut_flow}, we list all cuts in the first column. We 
display the total number of $(Z\rightarrow \nu\bar \nu)+j$ (second 
column), $W(\rightarrow \ell\nu_\ell)+j$ (third column), 
$W(\rightarrow\tau\nu_\tau)+j$ (fourth column) and $t\bar t$ events 
(fifth column) for an integrated luminosity of $\ifb{100}$ at the LHC 
with $\sqrt{s}=\tev{14}$. The signal $S$, the resulting ratio between 
signal and background ($B$) events and the estimate significance 
$S/\delta B$ are given in the sixth, seventh and eighth column, 
respectively.  
 
The significance of the signal depends on the error $\delta B$
(\ref{eq:error}) of the background.  In Section
\ref{subsec:backgrounds}, we discussed the individual systematical
errors. We also mentioned a data driven method to determine the
dominant $Z+j$ background from the $Z(\rightarrow \ell\ell)+j$
calibration channel. Our overall error estimate is then given by
\bea 
\delta B&=&\sqrt{5.3\,B_{Z+j}+ \sum_i B_{i}+\sum_i (0.1
  B_{i})^2},\label{eq:error}\\  
&&i=t\bar t\rm{, } \,W(\rightarrow \ell\nu_\ell)+j \rm{, } 
\,W(\rightarrow \tau \nu_\tau)+j \nonumber. 
\eea 

\begin{table*}[t!] 
\begin{center} 
\begin{ruledtabular} 
\begin{tabular}{c||c|c|c|c|c||c|c} 
cut & $Z(\rightarrow \nu\bar \nu)+j$ &  
$W(\rightarrow e\nu_e,\mu\nu_\mu)+j$ & 
$W(\rightarrow\tau\nu_\tau)+j$ & $t\bar t$  & signal $\:$ & $S/B$& 
$S/\delta B$ \\  
\hline 
$p_T(j_1)>\gev{500}$ & 27\,619 & 69\,802 & 35\,137 & 2\,206\,070 & 17\,797 &  
0.008 & 0.08\\\hline 
$\met>\gev{450}$ & 22\,798 & 20\,738 & 16\,835 &  63\,320 & 13\,350 &  
0.108 & 1.94\\ \hline 
veto on $e, \, \mu$ & 22\,284 & 6\,363 & 11\,978 & 23\,416 & 12\,810 & 
 0.200 & 4.68\\\hline 
veto on isolated taus & 22\,221 & 6\,274 & 9\,031 & 22\,848 & 12\,727 &  
0.21 & 4.96\\\hline 
veto on $b-$jets & 21\,295 & 5\,968 & 8\,617 & 11\,424 & 11\,064 &  
 0.23 & 6.94\\\hline 
veto on second jet & 15\,415 & 3\,702 &  5\,128 & 1\,408 & 5\,848 & 
0.23 & 8.17\\ 
$(p_T(j_2)\leq\gev{100})$&&&&&&&\\\hline 
\end{tabular} 
\end{ruledtabular} 
\caption{Cut flow for the benchmark scenario of 
  Sect.~\ref{sec:benchmark_sceanrio} at the LHC with $\sqrt{s}=14$ TeV and an 
  integrated luminosity of $100\,\rm{fb}^{-1}$. In the second
    last column, we present the ratio between signal and background
    number of events. In the last column, we estimate the significance
    via $\delta B$ given in Eq.~(\ref{eq:error}).
\label{tab:cut_flow}} 
\end{center} 
\end{table*} 
 
We start with a cut on the hardest jet (including $b-$tagged 
jets). After applying this cut, $t\bar t$ is the dominant background; 
it is two orders of magnitude larger than the signal and the remaining 
SM background, as can be seen in the first row of Table 
\ref{tab:cut_flow} and Fig.~\ref{fig:jetpt1_distribution}. Because of 
the large $t\bar t$ background, the signal significance is still very 
small. Note that for lower stop masses, a less stiff cut on the 
hardest jet would be slightly more efficient but we optimize our cuts 
for heavier stops since we would like to determine the discovery 
reach. 
 
The rather hard cut on the missing transverse energy strongly
suppresses the $W+j$ and $t\bar t$ backgrounds, but, coming after the
hard cut on the $p_T$ of the first jet, has little effect on the
signal and on the $Z+j$ background. This holds for relatively small
mass splittings between the lighter stop and the lightest
neutralino. In these scenarios (including our benchmark scenario), the
charm jets are rather soft, leading to large missing transverse
energy. As the mass splitting increases, the charm jets become harder
and are more often reconstructed, decreasing $\met$. We therefore
anticipate that the significance of our signal will be worse for
larger mass splittings (see below).
 
As we have shown in Fig.~\ref{fig:nlepton_distribution}, the lepton 
veto should efficiently reduce the SM background, while having little 
effect on the signal. We can see in Table \ref{tab:cut_flow} that the 
leptonic $W+j$ background is reduced by about a factor of three. $W+j$ 
events involving leptonically decaying taus from the $W$ are also 
removed. This cut also reduces the $t\bar t$ background 
significantly. Naively, one would assume that after demanding large 
missing transverse energy, at least one $W$ boson from $t\rightarrow 
b+W$ or in $W+j$ decays leptonically. However, there is a quite 
substantial probability that a charged lepton is not reconstructed 
according to the criteria described in Sect.~IIIB. Finally, the 
irreducible $Z+j$ background is not affected by this cut. 
 
The tau veto removes $25\%$ of the $W(\rightarrow \tau \nu)+j$ 
background events. However, nearly all $t\bar t$ events pass the 
cut. Requiring a large missing transverse energy cut and a lepton veto 
should mostly leave $t\bar t$ events with one $W$ decaying into a 
tau. Even so, only a few $t\bar t$ events are removed, since the 
$\tau$ tagging efficiency is very poor for $t\bar t$ events. 
 
After these four cuts, the signal significance is slightly less than 
five, with a signal to background ratio of 0.21. At this stage $t \bar 
t$ and $Z+j$ are still the dominant backgrounds. The $b-$jet veto 
further suppresses the $t \bar t$ background by a factor of 
two. As expected, it has little effect on the $Z+j$ and $W+j$ 
backgrounds. We saw in Fig.~4 that a relatively large fraction of the 
signal events contains a tagged $b-$jet. Thus the veto also removes 
$13\%$ of the signal events. Nevertheless this cut increases the 
signal significance to 6.94. 
 
The final cut vetoing a second hard jet is of crucial importance to 
further suppress the $t\bar t$ background. We now obtain a 
significance of $8.17$ and a rather good signal to background ratio of 
about $0.23$. Note that the $t \bar t$ background is now quite 
insignificant, being much smaller than the signal. It could be 
suppressed even further by reducing the $p_T$ threshold in the second 
jet veto. However, the number of signal events is decreased more 
strongly by this veto than the $Z+j$ background, such that our overall 
significance would get worse. 

\Mtext{
Searches for monojet signatures by the LHC experiments
\cite{ATLAS_monojet,CMS_monojet} also demand that the transverse momentum
vector of the second jet not be close to the missing $p_T$ vector
(ATlAS), or, equivalently, that the two hardest jets not be
back--to--back in the transverse plane (CMS). The ATLAS cut would
reduce our signal by about $35\%$, while the dominant $Z+$ jet
background would be reduced by only $12\%$. The reason is that in case
of the signal the visible stop decay products tend to be
back--to--back to the hardest jet, making it quite likely that the
second hardest jet is also approximately back--to--back with the
hardest jet. Such a cut is therefore not beneficial in our case.

The purpose of this cut is to suppress pure QCD multijet
backgrounds. However, ATLAS \cite{ATLAS_monojet} finds that even with
quite soft cuts on the hardest jet and the missing $E_T$, the pure QCD
background only amounts to about $5\%$ of the leading $Z+$ jet
background {\em without} such an angular cut. CMS \cite{CMS_monojet}
imposes stronger cuts on both the $p_T$ of the hardest jet and the
missing $E_T$. Before the angular cut, the multijet background is
about $20\%$ of the $Z+$jet background. Normally one expects the
importance of the pure QCD background to decrease with increasing
missing $E_T$ requirement. The fact that the QCD background before the
angular cut nevertheless is relatively more important in the CMS
analysis might be related to the fact that CMS counts muons as part of
the ``missing'' $E_T$. Semileptonic $c$ and $b$ decays can therefore
generate significantly larger missing $E_T$ as defined by CMS than in
the ATLAS definition. Our cut on missing $E_T$ is even harder than the
default cut used by CMS (200 GeV), for which the cut flow is shown; as
noted above, this should further reduce the relative size of the pure
QCD background. Moreover, the CMS ``multijet'' background is reduced
by about factor of 30 by a lepton veto, even though only isolated
leptons are vetoed. We conclude from this that our combination of a
very stiff missing $E_T$ cut with a lepton veto, which includes
non--isolated muons, should reduce the pure QCD background to a
negligible level even without an additional angular cut. 

We finally note that the Monte Carlo prediction of the total
background {\em exceeds} the background estimated by CMS directly from
data by about $25\%$ \cite{CMS_monojet}. This gives us some confidence
that our background estimate is conservative.}

\begin{figure}[t] 
\vspace{+1.5mm} 
\includegraphics[width=0.5\textwidth]{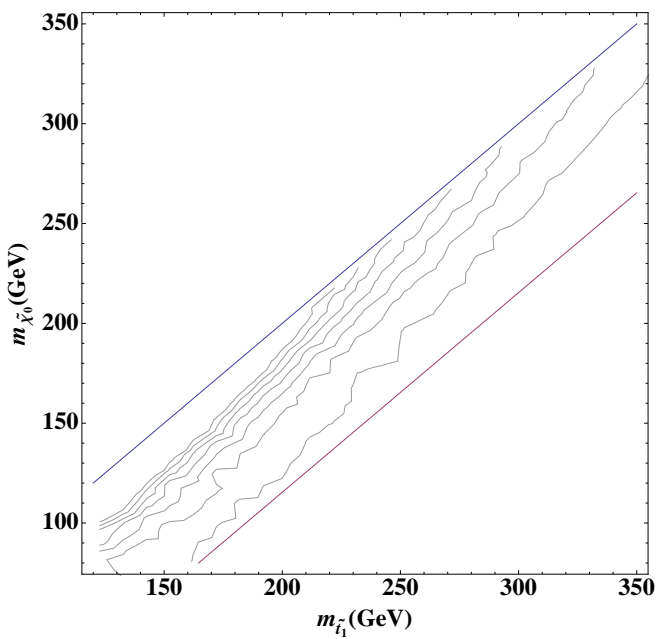} 
\caption{Number of signal events after cuts in the
    stop--neutralino mass plane assuming an integrated luminosity of
    $100\,\text{fb}^{-1}$ at $\sqrt{s}=14$ TeV. The two parallel
    straight lines delineate the region where $\tilde t_1 \rightarrow
    \tilde \chi_1^0 c$ decays are allowed but $\tilde t_1 \rightarrow
    \tilde \chi_1^0 W^+ \bar b$ decays are forbidden. The grey lines
    correspond to 7000, 6000, 5000, 4000, 3000, 2000 and 1000 signal
    events (from top to bottom), respectively.}
\label{fig:mstop_mneut_cross_distribution} 
\end{figure}

\begin{figure}[t] 
\vspace{+1.5mm} 
\includegraphics[width=0.5\textwidth]{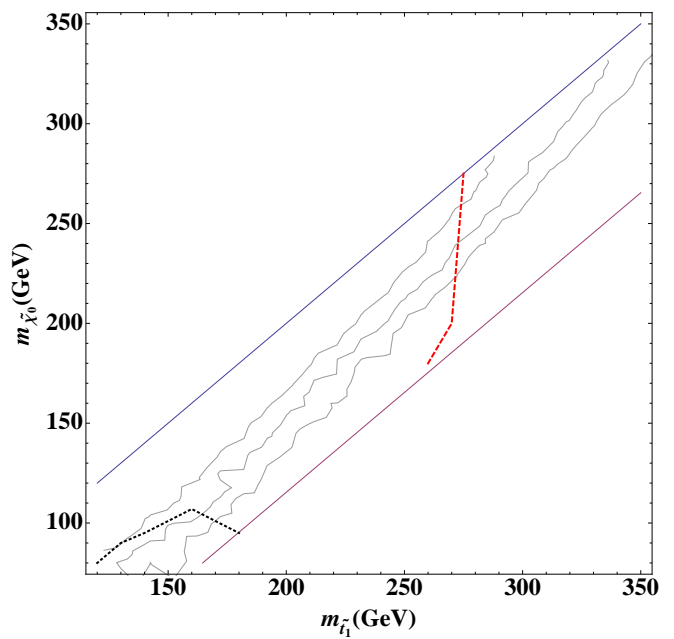} 
\caption{Signal significance with background error estimated as in
  Eq.~(\ref{eq:error}) in the stop--neutralino mass plane assuming an
  integrated luminosity of $100\,\text{fb}^{-1}$ at $\sqrt{s}=14$
  TeV. The two parallel straight lines delineate the region where
  $\tilde t_1 \rightarrow \tilde \chi_1^0 c$ decays are allowed but
  $\tilde t_1 \rightarrow \tilde \chi_1^0 W^+ \bar b$ decays are
  forbidden. The three grey lines correspond to 5$\sigma$, 3$\sigma$
  and 2$\sigma$ (from top to bottom), respectively. The short-dashed
  black curve delimits the Tevatron exclusion region, whereas the
  long-dashed red curve denotes the upper limit of the discovery reach
  of searches for light stops in events with two $b-$jets and large
  missing energy.}
\label{fig:mstop_mneut_distribution} 
\end{figure}

Having discussed the signal significance for our benchmark scenario, 
we now want to present results for other stop and neutralino 
masses. As before, we assume that all other sparticles are effectively 
decoupled. For the sake of simplicity, we apply the same cuts as for 
the benchmark point, {\it i.e.} the cuts in Table \ref{tab:cut_flow}. 
 
In Fig.~\ref{fig:mstop_mneut_cross_distribution}, we present
  the number of signal events in the stop--neutralino mass plane
  applying all cuts of Table \ref{tab:cut_flow}. The number of signal
  events is normalized to a luminosity of $\ifb{100}$ at
  $\sqrt{s}=\tev{14}$.
We see that even after the stiff cuts listed at the beginning
  of this Subsection, our ${\cal O}(\alpha_S^3)$ signal process yields
  in excess of $1000$ signal events out to quite large stop masses, as
  long as the mass splitting to the $\tilde \chi_1^0$ is small.

In Fig.~\ref{fig:mstop_mneut_distribution}, we show the statistical
significance in the stop--neutralino mass plane for an integrated
luminosity of $\ifb{100}$ at $\sqrt{s}=\tev{14}$.  We present the
discovery reach corresponding to $5\sigma$, $3\sigma$ and $2\sigma$,
respectively; the latter should be interpreted as the region that can
be excluded at 95\% c.l. if no signal is found. The stop can
dominantly decay into a charm and a neutralino for $m_{\neutr} +m_c <
m_{\st} < m_{\neutr} + m_W + m_b$, the area lying between the two
straight lines in Figs.~\ref{fig:mstop_mneut_distribution} and
  \ref{fig:mstop_mneut_cross_distribution}. The region below the
short--dashed black curve is excluded by Tevatron searches at the
$95\%$ confidence level \cite{Abazov:2008rc,CDFnote9834}. In the
region to the left of the long--dashed red curve, searches for light
stops in events with two $b-$jets and large missing energy
\cite{Bornhauser:2010mw} should have at least $5\sigma$ {\em
  statistical} significance.
 
We see from Fig.~\ref{fig:mstop_mneut_distribution} that the discovery 
of stop pairs in association with a jet should be possible for stop 
masses up to $\gev{290}$ and for mass splittings between stop and 
neutralino of up to $\gev{45}$. Stop masses up to $\gev{360}$ can be 
excluded at $2\sigma$ if the mass splitting is very small. As 
mentioned in the discussion of the missing $E_T$ cut as well as in 
Ref.~\cite{Carena:2008mj}, the significance of our monojet signal gets 
worse with increasing mass splitting. Increasing the mass splitting 
increases the average energy of the $c-$jets. This reduces the missing 
$E_T$, and increases the probability that the signal fails the veto on 
a second hard jet. These effects are cumulative: the reduced missing 
$E_T$ could be compensated by increasing the $p_T$ of the additional 
parton--level jet. However, this would also increase the $p_T$ of the 
$\tilde t_1 \tilde t_1^*$ pair, and hence the average $p_T$ of the 
$c-$jets from $\tilde t_1$ decay.  
 
The region close to the maximal allowed mass splitting (for the 
assumed loop--level two--body decay of $\tilde t_1$) could perhaps be 
probed through conventional searches for di--jet plus missing $E_T$ 
events, without demanding the presence of an additional parton--level 
jet. Alternatively one could reduce the missing $E_T$ cut, and try to 
suppress the $V+j$ backgrounds by a cut on the minimal multiplicity of 
charged particles \cite{Bornhauser:2010mw}. In both cases some sort of 
$c-$jet tagging would be helpful and perhaps even 
crucial. Unfortunately little is known (to us) about the capabilities 
of the LHC experiments to detect charm jets (or at least to isolate an 
event sample enriched in charm jets). We have therefore not attempted 
this approach here. 
 
The dashed red line in Fig.~\ref{fig:mstop_mneut_distribution} 
indicates that the two $b-$jet plus missing transverse energy 
signature \cite{Bornhauser:2010mw} is degraded less for larger mass 
splittings compared to our monojet signal. There the presence of two 
hard $b-$jets allowed to use a much milder missing $E_T$ cut of 
``only'' 200 GeV, and no veto against additional jet activity was 
used.  However, the analysis of \cite{Bornhauser:2010mw} isn't really 
comparable to our present work. First of all, only statistical 
uncertainties were considered in \cite{Bornhauser:2010mw}, whereas in 
the present case the uncertainty of the background, and hence the 
total significance, is dominated by the systematic errors; for 
example, after all cuts our benchmark point has a {\em statistical} 
significance of about $37$, compared to our stated significance of 
``only'' $8.17$. Secondly, detector effects were not included in 
\cite{Bornhauser:2010mw}. At least according to {\tt Delphes}, this 
over--estimates the efficiency of the lepton veto in reducing $W+j$ 
and top backgrounds. Finally, the red curve shown in 
Fig.~\ref{fig:mstop_mneut_distribution} holds under the assumption 
that there is a higgsino--like chargino just 20 GeV above the $\tilde 
t_1$; this increases the cross section for $\tilde t_1 \tilde t_1^* b 
\bar b$ production, which receives contributions from $\tilde t_1 
\tilde \chi_1^- \bar b$ production followed by $\tilde \chi_1^- 
\rightarrow \tilde t_1^* b$ decays (as well as charge conjugate 
processes). 
 
As noted above, the total uncertainty of our background estimate is
dominated by the systematic error on the $W + $1 jet background, which
we estimate to be 10\%. This is compatible with recent preliminary
ATLAS results on monojet searches at the 7 TeV LHC
\cite{ATLAS_monojet, CMS_monojet}. Since with the accumulation of
additional data our understanding of $W + 1$ jet production should
improve, we consider this estimate, and the resulting estimate of the
LHC reach, to be quite conservative. For example, ref.\cite{Vacavant}
estimates the {\em total} squared uncertainty from all $W, \, Z + 1$ jet
backgrounds to be $7 B_{Z+j}$. This would reduce the total uncertainty
$\delta B$ of the background after all cuts from about 715 (our
estimate) to about 360, i.e. by a factor of two. Once the total error
on the background has been established,
Fig.~\ref{fig:mstop_mneut_cross_distribution} can be used to determine
the region of parameter space that can be probed at a given
significance.

Finally, our estimate of the signal $S$ also has uncertainties. Since
we define the significance as $S / \delta B$ the systematic
(theoretical) uncertainty on $S$ will only change the signal reach
appreciably if the uncertainty is sizable. Since we are employing
leading order ${\cal O}(\alpha_S^3)$ expressions for the parton--level
signal cross section, NLO corrections might indeed be sizable. One
often attempts to estimate their magnitude by varying the
factorization and renormalization scales. For example, for $m_{\tilde
  t_1} = 120$ GeV, setting both of these scales equal to the stop
mass increases the parton--level cross section before cuts to 49 pb;
this is a factor 1.6 larger than the value of 31 pb we quote in
Table~\ref{tab:signal_xs}, which has been computed using the MadGraph
default scale choices. Unfortunately no NLO calculation of squark pair
production with radiation of an additional jet has been performed as yet.
All other theoretical uncertainties (due to details of the QCD shower
and fragmentation or the choice of parton distribution functions) are
significantly smaller than this estimate of the uncertainty due to NLO
corrections.

\section{Summary and Conclusion}\label{sec:summary} 
 
In this work, we considered light stops nearly degenerate with the 
lightest neutralino, with mass splitting of at most a few tens of 
GeV. Including CP phases in a sufficiently light electroweak 
gaugino--higgsino sector can then lead to scenarios consistent with 
electroweak baryogenesis and provide the right amount of dark 
matter. However, in such a scenario the direct production of a pair of 
light stops in stop pair production is difficult to detect at a hadron 
collider like the LHC since the decay products of the stops are quite 
soft. 
 
One solution is to examine stop pair production in association with 
two $b-$jets~\cite{Bornhauser:2010mw}. This could not only serve as a 
stop discovery channel, it could also be used to constrain Yukawa 
couplings of superparticles. The mixed QCD--EW production channels are 
sensitive to the $\st-\charg-b$ coupling. However, in order to 
determine the value of this coupling from future data, it is necessary 
to know the stop mass so that the QCD contribution can be subtracted. 
 
In Ref.~\cite{Carena:2008mj} a different process with negligible EW 
contributions was proposed: stop pair production in association with a 
hard jet.  In this paper we reanalyzed this process with some 
significant improvements. Firstly, we included the $t\bar t$ 
background, which had been neglected in previous works. Secondly, we 
simulated the signal and full SM background with the recent Monte 
Carlo simulations including a detector simulation.  Finally, we 
optimized the selection cuts, which was not done in 
Ref.~\cite{Carena:2008mj}. 
 
We discussed all the relevant collider observables for a specific 
benchmark point with $m_{\st}=\gev{220}$ and $m_{\neutr} = 
\gev{210}$. This led to a set of optimal cuts. We found that demanding 
a lot of missing transverse energy ($\met\ge\gev{450}$) and large 
transverse momentum of the hardest jet ($p_T(j_1)\ge\gev{500}$) is not 
sufficient to see an excess above the SM background. However, 
additionally imposing a lepton veto and a veto on the second jet 
($p_T(j_2)\le \gev{100}$) is very efficient for background 
suppression, the remaining dominant background process being the 
irreducible process $Z(\rightarrow \nu\bar\nu)+j$. Fortunately, this 
background can be determined experimentally from $Z(\rightarrow 
\ell\ell)+j$, although with reduced statistics. Here, we adopted a 
conservative estimate of the background uncertainty of the 
$Z(\rightarrow \nu\bar\nu)+j$ channel from Ref.~\cite{Vacavant} using 
$\delta B_{Z (\rightarrow \nu\bar\nu)+j} = 5.3 B_{Z(\rightarrow 
  \nu\bar\nu)+j}$. On the remaining SM backgrounds we assumed a 
systematic error of $10\%$. For our benchmark point, we showed that we 
can have a total signal significance exceeding $8$ for an integrated 
luminosity of $\ifb{100}$ at $\sqrt{s}=\tev{14}$. For the same cuts, 
we examined the discovery reach in the stop--neutralino mass plane and 
showed that this process can probe stop masses up to $\gev{290}$ if 
the mass splitting to the LSP is very small. This is well above the 
maximal $\tilde t_1$ mass compatible with electroweak baryogenesis in 
the MSSM. 
 
After having re--examined the monojet signature in this work, we are 
currently investigating the potential of reconstruction the $\st - 
\charg - b$ coupling by comparing the monojet channel to the two 
$b-$jets plus missing energy channel. Moreover, it would be 
interesting to devise methods to probe light top squarks with 
somewhat larger mass splitting to the LSP. Considering the importance 
that top squarks play in naturalness arguments for supersymmetry, 
this offers a good motivation to our experimental colleagues to 
investigate charm tagging at the LHC.

 
\begin{acknowledgments} 
  We thank S. Grab for useful discussions and A. Thomas for reading
  the manuscript. J.S.K. thanks the University of Bonn and the Bethe
  Center for hospitality during numerous visits. This work is
  supported in part by the Initiative and Networking Fund of the
  Helmholtz Association, contract HA-101 (``Physics at the
  Terascale''), by the Deutsche Telekom Stiftung, by the
  Bonn-Cologne Graduate School of Physics, by the German ministry for
  scientific research (BMBF) and by the ARC Centre of Excellence for
  Particle Physics at the Terascale.
 
\end{acknowledgments} 

\begin{appendix}
\section{Jet parton matching}\label{Matching}
\begin{table*}[t!]
\begin{center}
\begin{ruledtabular} 
\begin{tabular}{c||c|c|c}
sample & $p_T(j_1)>500$ GeV  & $\met>450$ GeV & jet veto\\
\hline
unmatched & 28048 & 23080 & 16002\\
matched & 63021 & 31478 & 16387
\end{tabular}
\caption{Cut flow for an unmatched $Z+\,1j$ sample and a matched $Z+\,1j$
  and $Z+\,2j$ sample. We assumed an integrated luminosity of 100
  fb$^{-1}$ at the LHC with $\sqrt{s}=14$ TeV.}
\label{table:matching}
\end{ruledtabular} 
\end{center}
\end{table*}
%

%

In our work, we have estimated the SM backgrounds to our monojet
signal by using the exact leading order matrix elements of the $V+1$
jet ($V=W,\,Z$) processes, where additional partons are emitted via
the parton shower. However, it is common lore that a parton shower is
only valid in the limit of soft and collinear emissions. Since we
apply a (relatively loose) jet veto on the second jet, it should be
investigated whether jet matching for the second jet affects our
results of the $V+\,$j backgrounds, see also the discussion in
Sect.~\ref{sec:Numerical_tools}.

Here, we shortly study the impact of the exact matrix element
calculation of single gauge boson production in association with up
two parton final states at the example of the $Z +$ jet background. We
use the parton-jet MLM matching algorithm \cite{mlm}, which is based
on event rejection. The numerical matching is performed with {\tt
  Madgraph5.1.4.6} interfaced with the shower generator {\tt
  Pythia6.4}, where we applied a $p_T$ sorted parton
shower. Hadronization is included, but we switch off the underlying
event for this study. (It has of course been included in the analysis
presented in the main text). We produce a matched sample of $Z +
1\,$jet and $Z + 2\, $jet events with subsequent $Z\rightarrow
\nu\bar\nu$ decays. We demand parton level cuts on the missing
transverse energy of $E_T>150$ GeV and minimum transverse energy cut
on the leading jet with $p_T>150$ GeV. We have chosen a merging scale
of $Q=60$ GeV and kept the remaining default settings in
Madgraph. \Mtext{We checked that this choice of matching scale leads
  to smooth distributions, and that our results are stable against
  small variations of the matching scale.} We compare the matched
sample with a description very similar to that used in the main text,
where $Z+1\,$parton events generated with exact matrix elements are
interfaced with {\tt Pythia6.4}. Here the second jet arises from the
parton shower.

Our numerical results are summarized in Table~\ref{table:matching}. We
present the number of events after applying the relevant kinematic
cuts, scaled to an integrated luminosity of 100 fb$^{-1}$ at
$\sqrt{s}=14$ GeV for both samples. We start with a $p_T$ cut on the
hardest jet. Matching increases the cross section after this first cut
by a factor $2.25$, as one can see in the second column. The first cut
shows that the parton shower description produces a less energetic
leading jet than the 2--jet matrix element description, although the
first jet is exact in both samples. The reason for this is as follows.
In the unmatched case, the second jet prefers phase space
configurations close to the leading jet. In the matched sample, the
second jet tends to be not as strongly centered around the leading jet
in phase space, and thus events where the second jet is closer to the
$Z$ increase the average $p_T$ of the leading jet. 

However, as a consequence of the second jet being closer to the $Z$ on
average, the missing transverse momentum distribution becomes softer
since the preferred kinematical configuration is determined by
minimizing the \Mtext{squared partonic center of mass energy} $\hat
s$. \Mtext{Thus matching reduces considerably the efficiency of the
  missing transverse momentum cut, so that the ratio of cross sections
  with and without matching is reduced to 1.36}. Finally, we find that
after \Mtext{imposing} the jet veto, both estimates of the
\Mtext{cross section} agree well because, as expected, the second jet
from the exact matrix element is on average harder than the parton
shower jet in the unmatched case. In other words, matching increases
the number of events with hard second jet, which are in any case
rejected by our jet veto. We conclude that the \Mtext{background
  estimates with and without MLM matching} agree well within the
statistical \Mtext{uncertainties}.
 
\end{appendix}

\bibliographystyle{h-physrev}

 
\end{document}